\documentstyle[pre,aps]{revtex}
\begin{document}
\draft
\font\sqi=cmssq8
\def\beq{\begin{equation}}
\def\eeq{\end{equation}}
\def\DR{\rm I\kern-1.45pt\rm R}
\def\DC{\kern2pt {\hbox{\sqi I}}\kern-4.2pt\rm C}
\def\DH{\rm I\kern-1.5pt\rm H\kern-1.5pt\rm I}
\newcommand{\bpi}{\mbox{\boldmath $\pi$}}
\def\theequation{\arabic{equation}}
\newcommand{\sgn}[1]{\;{\rm sgn}\; #1\;}
\def\G{{\cal G}}
\newcommand{\ket}[1]{\left|#1\right\rangle}
\newcommand{\bra}[1]{\left\langle #1\right|}
\newcommand{\cH}{{\cal H}}
\renewcommand{\imath}{{\rm i}}
\twocolumn[\hsize\textwidth\columnwidth\hsize\csname
@twocolumnfalse\endcsname

\title{Two phases of the  noncommutative quantum mechanics}
\author{Stefano Bellucci $^{1}$, Armen Nersessian $^{1,2,3}$
and Corneliu Sochichiu$^{2,4}$}
\address{$^1$INFN, Laboratori Nazionali di Frascati,
P.O. Box 13, I-00044, Frascati, Italy\\
$^2$JINR, Bogoliubov Laboratory of Theoretical Physics,
141980 Dubna, Russia\\
$^3$ Yerevan State University,
A. Manoogian St., 3, Yerevan, 375025 Armenia\\
$^4$ Institutul de Fizic\u a Aplicat\u a A\c S, str. Academiei, nr. 5, Chi\c sin\u
au, MD2028 Moldova}
\date{\today}
\maketitle
\begin{abstract}
We consider  quantum mechanics on the noncommutative plane in the presence of
magnetic field $B$. We show, that the model has two essentially different
phases separated by  the point $B\theta=c\hbar^2/e$, where $\theta$ is a
parameter of noncommutativity. In this point the system reduces to
exactly-solvable one-dimensional system. When $\kappa=1-eB\theta/c\hbar^2<0$
there is a finite number of states corresponding to the given value of the
angular momentum. In another phase, i.e. when $\kappa>0$ the number of
states is infinite. The perturbative spectrum near the critical point
$\kappa=0$ is computed.
\end{abstract}
  ]

% ----------------------------------------------------------------
\subsection*{Introduction} Recently some interest to
quantum mechanics on noncommutative space (noncommutative quantum mechanics)
arose, inspired by the development of string theory \cite{witten}. Beyond
the string theory meaning such model models also appear in various systems
describing spinning particles. They serve for the study of one-particle
sectors of noncommutative field theories arising from string considerations,
Quantum Hall effect, and general phenomenological impacts of the
noncommutativity \cite{Dunne:1990hv}--\cite{mp}. In particular in
refs.\cite{np,mp} noncommutative Landau problem on plane, sphere and torus
have been considered. A ``critical point'' was observed in these models when
the density of states becomes infinite (see also \cite{Duval:2000xr}). From
the algebraic point of view it corresponds to the degeneracy of the
representation of the Heisenberg algebra \cite{cornel1}.

% ----------------------------------------------------------------
The noncommutativity of coordinates is implemented by the relation,
\begin{equation}\label{nc}
[x^i,x^j]=i\theta^{ij},
\end{equation}
where $\theta^{ij}$ are $c$-numbers with the dimensionality
(length)$^{-2}$.

In the case when $[p_i,p_j]=0$, the noncommutative quantum mechanics reduces
to the usual one described by Schr\"odinger equation \cite{Mezincescu:2000zq}
\begin{equation}
\cH (p,{\tilde x})\Psi({\tilde x})=E \Psi({\tilde x}), \mbox{ where } {\tilde
x}^i=x^i-\frac 12 \theta^{ij}p_j.
\end{equation}
Hence, the difference between noncommutative and ordinary quantum mechanics
consists in the choice of polarisation only.

In this note we consider a two-dimensional noncommutative quantum mechanical
system with arbitrary central potential in the presence of constant magnetic
field $B$. It is given by the Hamiltonian,
 \beq {\hat{\cal
 H}}=\frac{{\bf p}^2}{2\mu}+ V(|{\bf x}|^2), \label{h0}
 \eeq
where the operators ${\bf p}, {\bf x}$ obey the commutation relations
 \begin{equation}\label{xp}
 [x^1,x^2]=i\theta,\quad [x^i,p_j]=i\hbar \delta^i_j,\quad
 [p_1,p_2]=i\frac{e}{c}B,
 \label{algebra}
 \end{equation}
we assume $\theta>0$.

In the absence of magnetic field, $B=0$, the leading part of Hamiltonian for
large noncommutativity parameter $\theta$ is given by the potential term,
while the remaining part can be considered as a perturbation \cite{gamboa1}.

In what follows we show, that in the presence of magnetic field
one has a parameter
 \beq
 \kappa=1-\frac{eB}{c\hbar^2}\theta.
 \eeq
which can be made small for arbitrary $\theta$, by choosing a proper value
of $B$. At the critical point,
 \beq
 \kappa=0,
 \eeq
the model becomes exactly solvable. This allows to develop the perturbative
analysis for the small $\kappa$ (and arbitrary $\theta$).

Surprisingly, it appears that the global properties of the model are
qualitatively different  for either $\kappa$ is positive or negative.
Although the perturbative analysis is applicable for both cases, for
negative $\kappa$  we can find some energy levels exactly.

% ----------------------------------------------------------------
\subsection*{The model}
Let us consider the system (\ref{h0},\ref{xp}) in more details.
 For this purpose
let us split the algebra (\ref{algebra}) in two independent
 subalgebras  and pass
to the operators $\pi_i$ and $x^i$ satisfying the following relations
 \beq
{\pi}_i=p_i-\frac{\hbar\varepsilon_{ij}{x^j}}{\theta}\;:\;
 [\pi_i,x^j]=0,\;\;  [\pi_1,\pi_2]=-i\frac{\hbar^2}{\theta}\kappa.
 \eeq
The $[x,x]$-commutator is given by Eq. (\ref{nc}).

  In  terms of these operators the Hamiltonian (\ref{h0}) reads
 \beq
 {\cal H}=\frac{{\bpi}^2}{2\mu}+
\hbar \frac{\pi_1x^2-\pi_2 x^1}{\mu\theta}
 + \hbar^2\frac{|{\bf x}|^2}{2\mu\theta^2}+
  V(|{\bf x}|^2).
 \label{8}\eeq

As one can see, there is a ``critical point'' for
 \beq
 \kappa=0
 \Leftrightarrow\; B={c\hbar^2}/{e\theta}.
 \eeq
At this point  operators $\pi_i$ belong to the center of quantum algebra,
and consequently, are constant ones. Thus, the system becomes effectively
one-dimensional. Also, from the requirement of rotational invariance it
follows that
 \beq
 \pi_i=0\;\Rightarrow\; {\cal H}=
 \frac{\hbar^2|{\bf x}|^2}{2\mu\theta^2}+{ V}(|{\bf x}|^2).
 \eeq
For this Hamiltonian it is easy  to find the exact energy
spectrum,
 \beq
 E^{(0)}_{n}=\frac{\hbar^2(n+1/2)}{\mu\theta}+V(\theta (2n+1)),
  \quad n=0,1,\ldots
 \label{E0}\eeq

Consider now the case of nonzero $\kappa$. In this case it is convenient to
introduce the creation and annihilation operators
 \beq
 { a}^{\pm}=\frac{x^1\mp \imath x^2}{\sqrt{2\theta}},
  \quad { b}^{\pm}=\frac{\sqrt{\theta}}{\hbar}\frac{\pi_1\mp \imath
 \pi_2}{\sqrt{2|{\kappa}|}}\;,
 \eeq
with the following non-zero commutators
 \beq
 [{ a}^-, { a}^+]=1,\quad
 [{ b}^-, { b}^+]= -\sgn \kappa.
 \eeq
In terms of these operators the Hamiltonian (\ref{8}) is of the form
 \begin{eqnarray}\nonumber
 {\cal H}=\frac{|\kappa|\hbar^2}{2\mu\theta}(b^+b^-+b^-b^+) -\\ \label{H}
 -\imath
 \frac{\sqrt{|\kappa|}\hbar^2}{\mu\theta}(b^+a^--a^+b^-)+\\ \nonumber
 +\frac{\hbar^2(a^+a^-+a^-a^+)}{2\mu\theta}+
  {V}(\theta(a^+a^- +a^-a^+)).
 \end{eqnarray}
The rotational  symmetry of the system corresponds to the
conserved angular momentum given by the operator,
 \beq
 2J=a^+a^--\sgn\kappa b^+b^-, \qquad [\cH,J]=0.
 \label{J}
 \eeq
As it can be seen, when  $\kappa<0$, the system is naturally formulated in
terms of representations
 of  the algebra $\G=su(2)$. For
$\kappa>0$, one has instead representations of $\G=su(1,1)$. The generators
of theses algebras are given by following operators,
 \beq
 L_\pm=b^\mp a^\pm,\qquad L_3=\frac{1}{2}(a^+a^-+\sgn\kappa b^+b^-).
 \eeq

It is worthwhile to note, that the angular  momentum of the system
given by (\ref{J}),  define the Casimir operator of the algebra
$\G$
\begin{equation}
J(J+\sgn\kappa)= L^2_3+\frac{\sgn\kappa}{2}(L_+L_-+L_-L_+).
\end{equation}
According to above, the Hilbert space splits in the irreducible
representations (irreps) of the algebra $\G$ which are
parameterized by the eigenvalues of $J$. Inside an irrep one can
introduce the basis labelled by the eigenvalue of $L_3$. As a
result we have the orthonormal basis in the Hilbert space
consisting of states
 \beq\label{basis}
 \ket{j,l}=\frac{(a^+)^{j+l}(b^+)^{j-l}}{\sqrt{(j+l)!(j-l)!}}\ket{0,0},
 \eeq
where $j$ and $l$ are eigenvalues of $J$ and $L_3$ respectively.
 Let us note that
the system of states  is equivalent to one of  a pair of coupled
oscillators. The angular momentum corresponds to the  total
occupation number
 \beq
 2j=n_a-\sgn\kappa n_b.
 \eeq

One can see that the  spectrum has different structure depending on the sign
of $\kappa$. Indeed, for $\kappa<0$ (or equivalently,
$B>{c\hbar^2}/{e\theta}$),
 the angular momentum $2j$ and the occupation number $n_a$
 corresponding to the operator $|{\bf
x}|^2/2\theta$, take the values
 \beq
 \begin{array}{l}
 n_a=0,1,\ldots, \\ 2j=n_a, n_a+1, \ldots\quad .
 \end{array}
 \eeq
For  $\kappa >0$ ($B<{c\hbar^2}/{e\theta}$) corresponding to the
non-compact case $\G=su(1,1)$, the eigenvalues of the angular
momentum $2j$ and of the operator $|{\bf x}|^2/2\theta$,
respectively take the values
 \beq
 \begin{array}{l}
 n_a=0,1,\ldots, \\
 2j=-\infty,\ldots, -1,0,1, \ldots, n_a.
 \end{array}
 \eeq

{\it Thus, the character of the spectrum
 essentially depends on the value of
magnetic field: for $B\theta<{c\hbar^2}/e$
 the angular momentum has the upper
bound, equal to the eigenvalue of the operator $|{\bf
x}|^2/2\theta$, while for $B\theta>{c\hbar^2}/e$ the eigenvalue of
$|{\bf x}|^2/2\theta$ becomes the lower bound for the angular
momentum.}

\subsection*{The spectrum}

In the basis (\ref{basis}) the Hamiltonian (\ref{H}) splits in the diagonal
part given by first and third lines and non-diagonal part given by the
second line. Let us consider the diagonal part given by the third line of
(\ref{H}) as the bare Hamiltonian. The remaining part can be considered as a
perturbation of the order $|\kappa|^{1/2}$. Then perturbation expansion
around the critical point $\kappa=0$ applies when
\begin{equation}
  \sqrt{|\kappa|j}\ll 1+\mu\theta V(n)/\hbar^2.
\end{equation}
The energy spectrum of the non-perturbed Hamiltonian is given by the
expression (\ref{E0}).

The first order correction to the $n-$th energy level vanishes, while the
computation of the second order correction yields the result
 \begin{eqnarray}\nonumber
& E_{(j,n)}^{pert}&=\frac{\kappa\hbar^2(2j-n)}{\mu\theta}
\left(1+\frac{n+1}{\Omega_{n+1}}-\frac{n}{\Omega_{n}}\right)-\\
&-&\frac{|\kappa|\hbar^2(n+1/2)}{\mu\theta\Omega_{n}}+\\
&+&\frac{\hbar^2(n+1/2)}{\mu\theta}+ V(2\theta n+\theta ), \nonumber
  \end{eqnarray}
where $$\Omega_{n}=\frac{\mu\theta}{\hbar^2} \left( V(2\theta
n+\theta)- V(2\theta n-\theta)+ \frac{\hbar^2}{\mu\theta}
\right).$$

% ----------------------------------------------------------------
Beyond this, in the compact case ($\G=su(2)$, $\kappa<0$), one can compute
exactly some energy levels in the ``lower'' (i.e. corresponding to small
$j$) part of the spectrum. In the mentioned case, the half-integer
eigenvalues $j$ and $l$ span the range $j=0,1/2,1,\dots$ and $-j\leq l \leq
j$. This happens due to finite dimensionality of the irreps of $\G$. The
Hamiltonian acts invariantly in each irrep because it commutes with the
Casimir operator $J$. Therefore, the problem of diagonalisation of the
Hamiltonian in the whole Hilbert space reduces to ``smaller'' problems of
diagonalisation in each finite-dimensional irrep.

Thus, the eigenvectors of $\cH$ can be represented as linear combination of
basis elements with the same number $j$,
 \beq\label{nj}
 \ket{j,s}=\sum_{l=-j}^{j}C_{l}^{(j,s)}\ket{j,l}, \qquad
 \cH\ket{j,s}=E_{(j,s)}\ket{j,s},
 \eeq
where $C_{l}^{(j,s)}=\langle{j,l}\ket{j,s}$, and half-integer number $s$
enumerates energy levels inside irrep.

The second equation in (\ref{nj}) can be rewritten as a set of $2j+1$ linear
equations for $C_l$ (we drop the superscripts $(j,s)$):
 \beq\label{hj}
 \begin{array}{c}
 (|\kappa| (j-l)+v(j+l)-\varepsilon ) C_l\\
 +\imath|\kappa|^\frac{1}{2}\left(\sqrt{(j-l+1)(j+l)}C_{l-1}\right.\\
 -\left.\sqrt{(j-l)(j+l+1)}C_{l+1}\right)=0,
 \end{array}
 \eeq
where we introduced shorthand notations for $\varepsilon$ and
$v(j+l)$ implicitly  defined by the equations
 $$
 E=\frac{\hbar^2}{\mu\theta}
\left(\varepsilon+\frac{1}{2}(1+|\kappa|)\right)+V(\theta)
 $$
and
 $$
 v(j+l)=j+l+\frac{\mu\theta}{\hbar^2}\bigl(
V(\theta(j+l+1))-V(\theta)\bigr).
 $$
This defines a $2j+1$ dimensional eigenvalue problem which can be solved by
standard linear algebra methods for not very large $j$, as well as
numerically if $j$ is large. In particular, for $j=0$ and $j=1/2$ the
corresponding energy levels are given by,
 \beq
 E_{(0,0)}= V(\theta)
\eeq and
\begin{eqnarray}
&&E_{(\frac{1}{2},\pm
\frac{1}{2})}=V(3\theta)+\frac{\hbar^2(1+|\kappa|)}{\mu\theta}\pm
\\ \nonumber
&\pm &
\left[{4|\kappa|}\left(\frac{\hbar^2}{\mu\theta}\right)^2+
\left((1-|\kappa|)\frac{\hbar^2}{\mu\theta}+
(V(3\theta)-V(\theta))\right)^2\right]^{1/2}
\end{eqnarray}
respectively.

Let us note, however, that the lowest  $j$ states  do not necessarily
correspond to the lowest energy levels. Depending on the form of the
potential, the higher $j$ states may have eigenvalues of the Hamiltonian
located in the lower part of the spectrum.

Unfortunately, in the case when $\kappa>0$ we cannot perform the same
analysis since in this case the representations of $su(1,1)$ are infinite
dimensional.

\subsection*{Example: Harmonic Oscillator}
Consider the particular case of harmonic oscillator,
\begin{equation}
V=\frac{\mu\omega^2|{\bf x}|^2}{2}.
 \end{equation}
 In this case one can solve the
spectrum exactly for any value of $\kappa$ \cite{np} (see also
\cite{Hatzinikitas:2001pm}). Our results agree with mentioned
ones. Let us  diagonalize the  Hamiltonian, performing the
appropriate (pseudo)unitary transformation:
\begin{equation}
 \left(\begin{array}{c}
 {a}\\
 {b}
 \end{array}\right)\to U\cdot
 \left(\begin{array}{c}
 a\\
 b
 \end{array}\right),
\end{equation}
where the matrix $U$ belong to SU(1,1) for $\kappa>0$ and to
SU(2) for $\kappa<0$. Explicitly,
\begin{equation}\label{u}
  U=\left\{
\begin{array}{c}
  \left(\matrix{
  \cosh \chi e^{\imath \pi/4}& \sinh\chi e^{\imath \pi/4}\cr
  \sinh\chi e^{-\imath \pi/4}&\cosh \chi e^{-\imath \pi/4}
  }\right),\quad \mbox{for } \kappa>0 \\ {}\\
  \left(\matrix{
  \cos \chi e^{\imath \pi/4}& \sin\chi e^{\imath \pi/4}\cr
  -\sin\chi e^{-\imath \pi/4}&\cos \chi e^{-\imath \pi/4}
  }\right),\quad \mbox{for } \kappa<0,
\end{array}
  \right.
\end{equation}
where ``angle'' $\chi$ is given by the following relations,
\begin{equation}
  2\chi=\left\{
  \begin{array}{c}
  \tanh^{-1}(2\sqrt{\kappa}/({\cal E}+\kappa)),\quad \mbox{for }
  \kappa>0\\
  \tan^{-1}(2\sqrt{-\kappa}/({\cal E}+\kappa)),\quad \mbox{for }
  \kappa<0
  \end{array}\right. .
\end{equation}
We have used here the notation ${\cal
E}=1+\left({\mu\omega\theta}/{\hbar}\right)^2$.

The diagonalized Hamiltonian reads,
\begin{equation}\label{osc}
 {\cal H}_{\rm osc}=
 \frac{1}{2}\hbar\omega_+(b^+b^- +b^-b^+)+\frac{1}{2}\hbar\omega_-(a^+a^-+
 a^-a^+),
\end{equation}
where
\begin{equation}
 \frac{2\mu\theta\omega_\pm}{\hbar}=\left\{
 \begin{array}{c}
  \pm\frac{1}{2}({\cal E}-\kappa)+\frac{1}{2}\sqrt{({\cal E}+\kappa)^2-4\kappa}\\
  \frac{1}{2}({\cal E}-\kappa)\pm
  \frac{1}{2}\sqrt{({\cal E}+\kappa)^2-4\kappa}.
 \end{array}\right..
\end{equation}

Hence, the spectrum  is of the form
\begin{equation}\label{oscspectrum}
 E^{\rm osc}_{n_1,n_2}=
 \hbar\omega_+\left(n_1+\frac{1}{2}\right)
 +\hbar\omega_-\left(n_2+\frac{1}{2}\right).
\end{equation}

Let us recall that the transformation (\ref{u}) belongs to the
symmetry group of the rotational momentum $J$. Therefore in new
variables its eigenvalues are given by the following equation,
\begin{equation}
j=n_1-\sgn\kappa n_2,
\end{equation}
where $n_1,n_2=0,1,2,\dots$ .

In particular, the vacuum energy corresponding to different signs
of $\kappa$ looks as follows,
$$h_0=\frac{\hbar}{\mu\theta}\sqrt{({\cal E}-\kappa)^2+4\kappa({\cal E}-1)},\quad
\mbox{for }\kappa>0,$$ and
$$h_0=\frac{\hbar}{\mu\theta}\left({\cal E}-\kappa\right),\quad \mbox{for
}\kappa<0.$$

\subsection*{Concluding Remarks}
In this paper we considered a two-dimensional central symmetric
noncommutative mechanical model in the presence of magnetic
field. We have shown that in the case when magnetic field is
smaller than some critical value the spectrum of the model is
organized according to representations of algebra su(1,1) while
for the magnetic field beyond this value it ``reorganizes''
according to representations of su(2). This algebras are symmetry
algebras of the rotational momentum operator in these two cases.
These cases are physically different. In particular in the first
one the possible values of rotational momentum span both positive
and negative half-integer numbers while in the second case only
positive orbital numbers are allowed. This may lead to the
conclusion that in the presence  of a strong magnetic field
properly oriented with respect to inverse noncommutativity
parameter $\theta^{-1}$ the spinning properties of the
noncommutative particle are gravely affected.

As an example we considered the particular case of the harmonic
oscillator. Our results appear to be in agreement with the ones
previously known in the literature.
% ----------------------------------------------------------------
\subsection*{Acknowledgments.} We thank Ph. Pouliot for  criticism
leading, hopefully, to the improving of the manuscript and useful
discussions on the oscillator example. A.N. thanks INFN for the
financial support and hospitality during his stay in Frascati,
where this work was started. A.N. and C.S. were partially
supported under the  INTAS project 00-00262. C.S. was also
supported by RFBR grants: \#99-01-00190 and young scientists
support grant, Scientific School support grant \# 00-15-96046.

\end{document}